\documentclass[aps,prl,twocolumn,superscriptaddress]{revtex4}
\usepackage{graphicx}
\usepackage{color}
\usepackage[hypertex,unicode=true,breaklinks=true]{hyperref}

\hypersetup{
        colorlinks=false,
        citecolor=blue,
        linkcolor=black,
        urlcolor=blue}
\renewcommand{\href}[2]{#2}

\newcommand{\halfl}{\ensuremath{{\scriptstyle \frac{1}{2}}}}

\newcommand{\un}   [1]{\ensuremath{\,\mathrm{#1}}}

\newcommand{\ket}  [1]{\ensuremath{\left | #1 \right \rangle }}

\newcommand{\Vdc} {\ensuremath {V_{\mathrm{DC}}}}

\newcommand{\sub}[1] {({#1})}
\newcommand{\subf}[1] {({#1})}

\begin{document}
\title{Classical non-Gaussian state preparation through squeezing in an opto-electromechanical resonator}

\author{M.~Poot}
\email{menno.poot@yale.edu} \affiliation{Department of
Electrical Engineering, Yale University, New Haven, CT 06520,
USA}

\author{K.~Y.~Fong}
\affiliation{Department of Electrical Engineering, Yale
University, New Haven, CT 06520, USA}

\author{H.~X.~Tang}
\email{hong.tang@yale.edu} \affiliation{Department of
Electrical Engineering, Yale University, New Haven, CT 06520,
USA}

\begin{abstract}
We demonstrate squeezing of a strongly interacting
opto-electromechanical system using a parametric drive. By
employing real-time feedback on the phase of the pump at twice
the resonance frequency the thermo-mechanical noise is squeezed
beyond the 3 dB instability limit. Surprisingly, this method
can also be used to generate highly nonlinear states. We show
that using the parametric drive with feedback on, classical
number-like and cat-like states can be prepared. This presents
a valuable electro-optomechanical state-preparation protocol
that is extendable to quantum regime.
\end{abstract}

\date{\today}

\maketitle

\section{Introduction}
The field of nano- and optomechanics has recently moved into
the domain of quantum \emph{mechanics}
\cite{poot_physrep_quantum_regime} by cooling resonators to the
ground state \cite{oconnell_nature_quantum_piezo_resonator,
chan_nature_groundstate, teufel_nature_groundstate}, achieving
strong coupling between photons and phonons
\cite{groeblacher_nature_strong_coupling,
weis_science_transparency, teufel_nature_strong_coupling}, and,
most recently, by entangling them
\cite{palomaki_science_entanglement}. Also, squeezing of light
mediated through optomechanical backaction has been
demonstrated \cite{brooks_nature_atom_motion_light_squeezing,
safavi-naeini_nature_optomechanical_squeezing,
purdy_PRX_optomechanical_squeezing}. Along the same lines,
experiments with quantum protocols are currently being explored
in the classical regime
\cite{palomaki_nature_coherent_transfer,
okamoto_natphys_coupled, faust_natphys_coherent_control}, but
first nonclassical states of the harmonic oscillator
\cite{bose_PRA_nonclassical_preparation,
vanner_PNAS_pulsed_optomechanics, vanner_PRX_linear_quadratic,
romero-isart_PRL_superpostions} will need to be prepared. As we
will demonstrate here experimentally, parametric squeezing is
well suited for this purpose when combined with real-time
control.

Squeezing is a powerful technique where the stochastic motion
of a resonator is reduced in one quadrature at the expense of
an increase in the other one. It is an important concept in the
context of backaction evading \cite{clerk_NJP_BAE_squeezing}
and pulsed measurements
\cite{vanner_PNAS_pulsed_optomechanics}. However, for the most
often used squeezing method where the resonance frequency
$\omega_0$ is modulated at twice that frequency
\cite{rugar_PRL_squeezing} (so-called parametric squeezing) the
maximum squeezing that can be achieved is only 3 dB; this
happens when the modulation amplitude of $\omega_0$ equals the
damping rate $\gamma_0$. Beyond this limit the other quadrature
of the resonator becomes unstable leading to regenerative
parameteric oscillations. Currently a number of ways to
circumvent this problem are being studied
\cite{vinante_PRL_feedback_squeezing,
szorkovszky_PRL_squeezing_weak_measurement}. Here we
demonstrate a feedback technique where the phase of the
parametric modulation is adjusted in real time and the 3 dB
limit is overcome. The method also naturally leads to
non-Gaussian states of the mechanical resonator.

\begin{figure}[bt]
\includegraphics{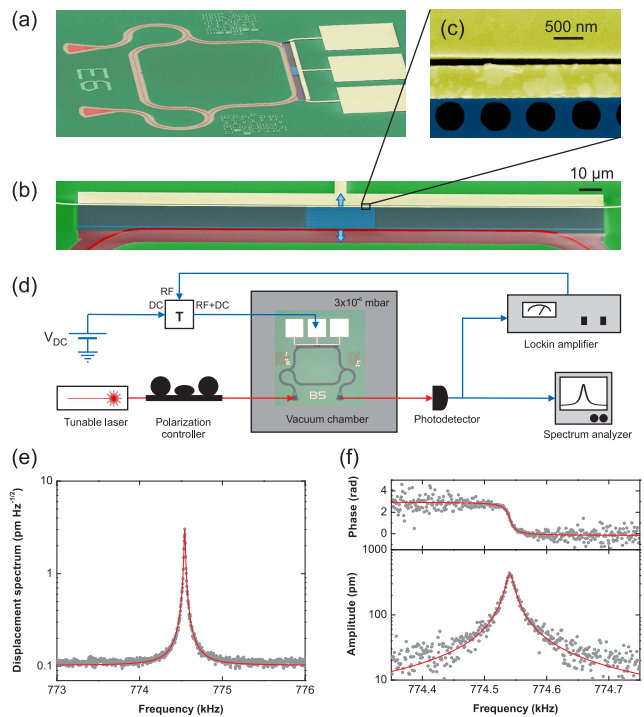}%
\caption{
\subf{a} Colorized scanning electron micrograph showing an overview of a
device taken under an angle; zooms of typical H-resonators are
shown in \sub{b} and \sub{c}. The waveguide is shown in red, the electrodes in yellow,
and the part of the resonator not covered by the moveable electrode in blue.
The arrows indicate the direction of motion for the fundamental in-plane mode.
\subf{d} The measurement setup.
\subf{e} Power spectral density of the displacement noise (symbols)
with fit (line) measured at $\Vdc = 0.5 \un{V}$. \subf{f} Driven
response at $\Vdc = 0.5 \un{V}$ and $V_d = 11 \un{\mu V}$. The
fit yields a quality factor of $62,000 \pm 1,000$.\label{fig:overview}}
\end{figure}

\section{The opto-electromechancial system and parametric squeezing}
To achieve a high degree of squeezing in an optomechanical
system, one wants to efficiently modulate the spring constant
$k$. Although optical readout can give unsurpassed displacement
sensitivity \cite{poot_physrep_quantum_regime}, typically
optical backaction effects are weak. Therefore, we combine
optical readout with strong electrostatic forces in an
integrated opto-electromechanical device (Fig.
\ref{fig:overview}\sub{a}-\sub{c}). Recently we used similar
nanofabricated devices as optomechanical phase-shifters
\cite{poot_apl_phaseshifter} and showed that these display very
strong electrostatic interactions \cite{poot_H_electrostatic}.
The movable part of the device consists of four thin arms that
connect to a rectangular block containing a photonic crystal.
An electrode runs over one pair of arms and is close to
another, fixed, electrode. The other side of the resonator runs
close to the waveguide of an on-chip Mach-Zehnder
interferometer (Fig. \ref{fig:overview}\sub{a}), enabling
sensitive displacement detection using the measurement setup
shown in Fig. \ref{fig:overview}\sub{d}. The fundamental
in-plane eigenmode of this ``H-resonator'' has a resonance
frequency around 723 kHz and the thermo-mechanical noise at
room-temperature is resolved with a signal-to-background ratio
of 29 dB as shown by the spectrum in Fig.
\ref{fig:overview}\sub{e}. By applying a static ($\Vdc$) and an
a.c. voltage ($V_d$) between the electrodes, the resonator is
actuated; Fig. \ref{fig:overview}\sub{f} shows the driven
response. The small driving voltage that is used is a
manifestation of the large electrostatic interactions in the
H-resonator.

In such strongly interacting electromechanical systems, the
resonance frequency can be tuned via the electrostatic spring
effect \cite{poot_H_electrostatic,
unterreithmeier_nature_dielectric,
kozinsky_APL_dynamicrange_tuning}. By recording driven
measurements (cf. Fig. \ref{fig:overview}\sub{f}) while
sweeping $\Vdc$, a curvature of $f''_0 = -2.0 \un{kHz/V^2}$ is
found. Hence, by applying a pump voltage $V_P \sin(2\omega_F t
+ \theta)$ the resonance frequency is modulated at twice the
reference frequency $\omega_F \approx \omega_0$ with an
amplitude $\chi = f''_0 \Vdc V_P$. Such a signal is called a
``2f'' parametric pump and its effect is most easily analyzed
using the complex amplitude of the resonator $A$. In the frame
rotating at $\omega_F$ it is defined as
\cite{poot_PRA_backaction_limits}:
\begin{equation}
A \equiv \left(u + \frac{\dot u}{i\omega_F}\right)\exp(-i\omega_F t). \label{eq:Adef}
\end{equation}
This means that an oscillating displacement $u(t) = A_0
\cos(\omega_F t)$ has a complex amplitude $A = A_0$ and
likewise $u(t) = A_0 \sin(\omega_F t)$ becomes $A = iA_0$. By
taking the time derivative of Eq. (\ref{eq:Adef}) and inserting
the equation of motion $m\ddot u = -k(t)u - m\omega_0\gamma_0
\dot u + F$ into that expression, the differential equation for
the dynamics of the complex amplitude is obtained. In the
rotating wave approximation (RWA) it is
\cite{rugar_PRL_squeezing}:
\begin{equation}
\dot A \approx i \Delta_0 A - \frac{\gamma_0}{2} A +
\frac{\chi}{2}e^{i\theta} A^* - \frac{i}{2} f_F, \label{eq:Adot}
\end{equation}
where $\Delta_0 = (\omega_0^2 - \omega_F^2)/2\omega_F \approx
\omega_0 - \omega_F$ is the detuning, $m f_F = \left\langle 2F
e^{-i\omega_F t}\right\rangle_{\text{RWA}}$ is the force acting
on the resonator in the RWA, and $m$ is its mass. The term with
$A^*$ in Eq. (\ref{eq:Adot}) indicates that the parametric pump
breaks the time invariance and leads to squeezing
\cite{rugar_PRL_squeezing}.

\begin{figure}[tb]
\includegraphics{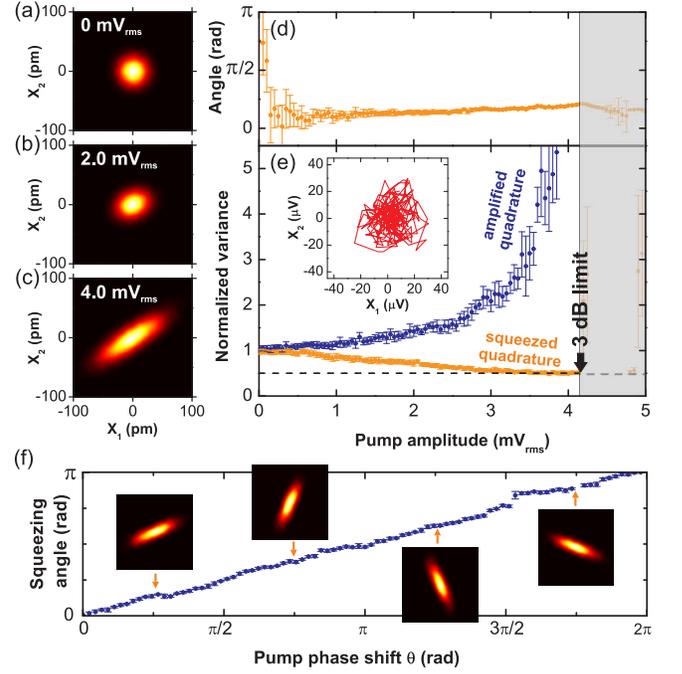}%
\caption{
Parametric squeezing of the thermal motion. \subf{a}-\subf{c} Colorplots of the measured pdf at the indicated pump power.
\subf{d},\subf{e} Evolution of the angle \sub{d} and variance
\sub{e} of the squeezing for increasing pump power. The
dashed line shows the 3 dB limit for parametric squeezing.
\subf{f} Squeezing angle versus the applied phase shift of the
2f drive signal. The insets show the pdf near $\theta = \pi/4,
3\pi/4, 5\pi/4, 7\pi/4$ respectively. All markers and error
bars in this figure are the mean value and standard deviation
of 10 individual measurements, respectively. The pdfs are
reconstructed from these 10 traces combined, and $\Delta_0 \approx 0$ everywhere.
\label{fig:parametric}}
\end{figure}

The rotating frame is not only a convenient representation for
the equation of motion, but is also directly accessible in the
experiment: a lockin amplifier (Zurich Instruments HF2; Fig
\ref{fig:overview}\sub{d}) demodulates the photodetector signal
at $\omega_F$ and, after scaling by the transduction factor,
its two quadratures $X_1$ and $X_2$ are the real and imaginary
part of $A$. The inset of Fig. \ref{fig:parametric}\sub{e}
shows a representative phase-space trajectory of such a
demodulated timetrace. Note that when the signal-to-noise ratio
between the mechanical signal and the imprecision noise is high
(see Fig. \ref{fig:overview}\sub{e}) the trajectories
accurately represent the mechanical motion. By repeatedly
measuring these trajectories, the probability density function
(pdf) of the resonator, which is the classical analogue of the
quantum Wigner function, can be reconstructed
\cite{note:pdf_resonstruction}. As shown in Fig.
\ref{fig:parametric}\sub{a} the Brownian motion of the
resonator appears as a circular Gaussian in the pdf in the
absence of a pump voltage, i.e. at $V_P = 0$.

By increasing $V_P$ the pdfs becomes ellipsoidal as fluctuation
in one quadrature decrease, whereas they are enhanced in the
orthogonal direction [Fig. \ref{fig:parametric}\sub{b,c}]
thereby squeezing the thermal motion. This can be quantified by
extracting the variances in these two directions (i.e., the
squared lengths of the minor and major axes of the ellipses).
As shown in Fig. \ref{fig:parametric}\sub{e} the squeezing
increases with increasing $V_P$ but when the normalized
variance of the squeezed quadrature reaches $\halfl$ the
fluctuations in the anti-squeezed quadrature grow exponentially
until limited by nonlinearities in the resonator
\cite{lifshitz_coupled_NL, karabalin_APL_parametric_VHF,
chung-chiang_APL_parametric_SWCNT}. Above this threshold $\chi
> \gamma_0$ and parametric oscillations result, which are
characterized by a large mean value of $A$. In our strongly
coupled high-quality electromechanical resonator
\cite{poot_H_electrostatic} this happens at $V_P = 4.1
\un{mV_{rms}}$ which is orders of magnitude lower than
previously reported thresholds for top-down devices
\cite{rugar_PRL_squeezing, mahboob_natnano_bit} and is even
below that of extremely floppy bottom-up devices
\cite{eichler_NL_parametric_amplification_CNT}. Interestingly,
the low threshold could be used to efficiently encode
information in the phase of the oscillations
\cite{mahboob_natnano_bit}, but here we focus on the
sub-threshold behavior. In that regime the maximum attainable
squeezing is thus limited to 3 dB.

\section{Overcoming the 3d limit using realtime pump-phase feedback}
The angle of the squeezing ellipse is set by the phase between
the 2f pump and the reference frame. Figure
\ref{fig:parametric}\sub{d} indicates that the squeezing angle
remains constant for fixed $\theta$. Changing the phase of the
reference frame rotates the apparent squeezing angle
\cite{rugar_PRL_squeezing}, but nothing happens to the actual
motion. However, when the \emph{pump phase} is varied, the
squeezing angle of the actual motion rotates. This is an
important distinction that enables squeezing beyond 3 dB as we
will show later. Figure \ref{fig:parametric}\sub{f} shows that
by adjusting $\theta$ the ellipsoidal pdf's rotate
proportionally to $\halfl \theta$.

\begin{figure}[tb]
\includegraphics{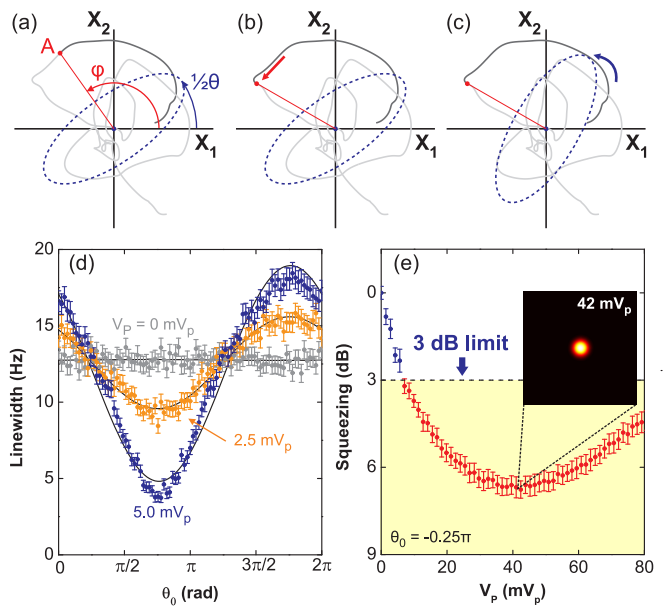}%
\caption{\subf{a-c}
Schematic illustrations of the real-time squeezing feedback process. The complex amplitude of
the resonator follows at trajectory in the $(X_1, X_2)$-plane
(gray). The red stick indicates the position at a certain time
\sub{a} and the ellipse indicates the squeezing direction. The
maximum squeezing and anti-squeezing occurs along the minor and
major axis respectively. \subf{b} Some time later the complex
amplitude has changed and no longer points along the minor axis
where the optimal squeezing occurs. By measuring the phase of the
complex amplitude, the squeezing angle can be corrected in
real-time \sub{c}.
\subf{d} Dependence of the linewidth on the
pump phase and pump power. Fits (solid lines) give $\chi = 0.03\pm 0.06, 3.01\pm 0.06, \text{~and~}
7.06\pm 0.13 \un{Hz}$ for $V_P = 0, 2.5, \text{~and~} 5.0
\un{mV}$ respectively. \subf{e} Squeezing below (blue) and beyond (red) the
3 dB limit. The inset shows the stationary pdf at the maximum squeezing
on the same scale as Fig. \ref{fig:parametric}\subf{a}.
All errorbars indicate fit uncertainties.
\label{fig:feedback}}
\end{figure}

In principle, squeezing exceeding 3 dB is possible when the
pump exceeds the threshold (i.e. $\chi
> \gamma_0$), but in this case the resonator will ring up to large
oscillation amplitudes and the squeezing cannot be stationary
\cite{blencowe_physb_squeezing}. By using a pump that is not
exactly at twice the resonance frequency (i.e. when $\Delta_0
\neq 0$) the threshold pump power can be increased, but still
the maximum amount of squeezing that can be obtained is limited
to 3 dB, although estimation schemes can further reduce the
uncertainty in the position
\cite{szorkovszky_NJP_squeezing_position_estimation}. Here we
use a different method, where the phase of the pump is adjusted
in real time \cite{gieseler_PRL_nanoparticle}, based on the
measured location of the resonator in phase-space as
illustrated in Fig. \ref{fig:feedback}\sub{a-c}. In particular,
we estimate $\varphi = \angle A$ using the phase $\hat \varphi$
measured by the lockin amplifier and use its programmable
digital-signal processor to update the 2f phase $\theta$ every
$70 \un{\mu s}$. When this adjustment is chosen carefully, the
squeezing direction can be optimized in real time as
illustrated in Fig. \ref{fig:feedback}\sub{a-c}. Equation.
(\ref{eq:Adot}) shows that when $\theta \rightarrow \theta_0 +
2\varphi(t)$, the dynamics of $A$ become independent of $A^*$.
However, in reality the actual phase $\varphi$ is not known and
only the estimated phase $\hat \varphi$ can be used. Inserting
this into Eq. (\ref{eq:Adot}) yields:
\begin{equation}
\dot A \approx \left(i \Delta_0 - \frac{\gamma_0}{2}  +
\frac{\chi}{2}e^{i\theta_0+2i(\hat\varphi-\varphi)}\right) A - \frac{i}{2} f_F, \label{eq:Adotfeedback}
\end{equation}
which no longer contains $A^*$. Equation
(\ref{eq:Adotfeedback}) also shows that when the phase estimate
is accurate (i.e. when $\hat\varphi =\varphi$) the resonance
frequency and damping rate can be adjusted through the phase
offset $\theta_0$ as $\Delta = \Delta_0 + \chi
\sin(\theta_0)/2$ and $\gamma = \gamma_0 - \chi \cos(\theta_0)$
respectively. Figure \ref{fig:feedback}\sub{d} shows the
dependence of the linewidth $\gamma$ on $\theta_0$ for three
different pump powers. Without a pump the linewidth remains
constant at $12.8 \un{Hz}$; with the pump on the linewidth
shows the sinusoidal dependence $\gamma = \gamma_0 + \chi
\cos(\theta_0 - \alpha)$ expected from Eq.
(\ref{eq:Adotfeedback}) ($\alpha$ is an offset due to delays in
the system). The maximum reduction in the thermal noise
coincides with the largest damping \cite{poot_APL_cooling},
i.e. at $\theta_0 = \alpha \approx -0.25\pi$. This value is set
and the pump power is increased. This squeezes the thermal
motion of the resonator as shown in Fig.
\ref{fig:feedback}\sub{e}. For low $V_P$ the amount of
squeezing increases with increasing pump power, just as in the
case without real-time feedback (cf. Fig.
\ref{fig:parametric}\sub{e}). The squeezing approaches 3 dB
around $V_P = 6.5 \un{mV_p}$, but now when increasing the power
further no instability is encountered and the motion is
squeezed beyond the 3 dB limit (dashed line). At $V_P = 42
\un{mV_p}$ the maximum squeezing of $6.7\pm 0.3 \un{dB}$ is
obtained. Further increasing the power reduces the squeezing
again.

This degradation is analogous to the process in active feedback
cooling where imprecision noise reheats the resonator at strong
feedback \cite{poot_physrep_quantum_regime,
kleckner_nature_feedback, poggio_PRL_feedback,
poot_APL_cooling}. In our case the imprecision noise makes that
the pump phase is not exactly at the optimal value $\alpha +
2\varphi - \pi/2$ as $\hat \varphi$ is not an accurate
estimation of $\varphi$. By improving the measurement
sensitivity the phase estimation will become better and the
maximum squeezing is enhanced. However, ultimately the
squeezing must be bounded by the zero-point motion
$u_{\mathrm{zpm}}$ due to the Heisenberg uncertainty principle
which requires $\Delta X_1 \cdot \Delta X_2 \ge
u_{\mathrm{zpm}}^2$. Since the feedback tracks $\varphi$ the
pdfs remain circular as shown in Fig.
\ref{fig:feedback}\sub{e}. This means that $\Delta X_1 = \Delta
X_2$, thus ultimately limiting the squeezing to $\Delta X_1 =
\Delta X_2 = u_{\mathrm{zpm}}$. However, the current device is
still far from this limit (which requires $\sim 72 \un{dB}$ of
thermomechanical noise squeezing at room temperature) but by
cooling the device with a dilution refrigerator this number can
be reduced by more than $40 \un{dB}$.

\section{Nonlinear feedback: classical non-Gaussian state generation}
\begin{figure}[tb]
\includegraphics{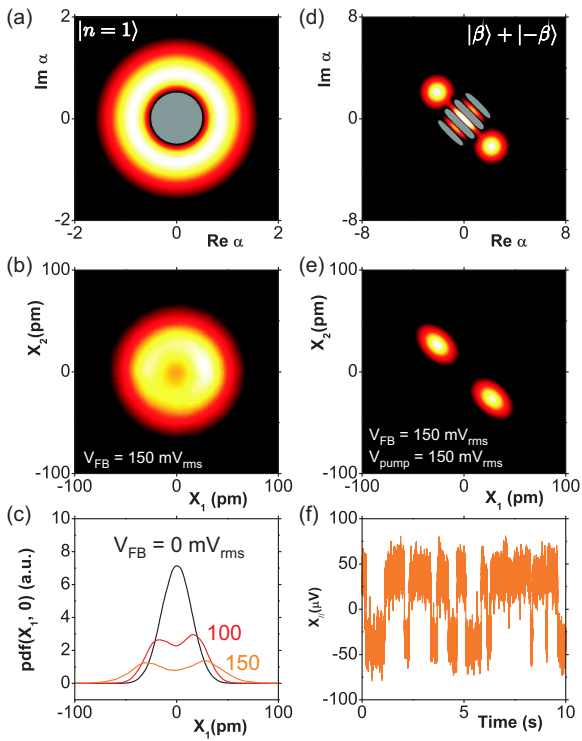}%
\caption{The calculated Wigner functions (top panels) for a quantum mechanical number
state $\ket{n=1}$ \sub{a} and a cat state $\ket{\beta} + \ket{-\beta}$ that consists
of a superposition of two coherent states with amplitude $\beta = 3 \exp(i\pi/4)$ \sub{d}.
The gray areas indicate regions where the Wigner function is negative.
In the classical regime these areas are washed out and the distribution
is always positive.
\subf{b},\subf{e} experimentally observed pdf's as prepared using our parametric feedback scheme.
\subf{e} horizontal cuts through pdfs measured for different
strengths of the 2f feedback.
\subf{f} measured dynamics of the quadrature along the long axis $X_{//}$ of the cat-like state shown in \sub{e}.
\label{fig:nongaussian}}
\end{figure}

Squeezing from all directions simultaneously (i.e.
isotropically) can be viewed as cooling, but we emphasize that
the mechanism that causes the reduction of the occupied phase
space in our parametric pump scheme with phase estimation is
very different from other, linear, cooling techniques.
Interestingly, our technique is inherently nonlinear and
naturally generates non-Gaussian classical states that do not
arise when simply cooling or parametrically driving the
resonator. Figures \ref{fig:nongaussian}\sub{b,e} show two
examples: a number-like and a cat-like state; the Wigner
functions of the corresponding quantum states are shown in
\sub{a} and \sub{d}. The Wigner function of the number state
$\ket{n=1}$ consists of a donut-shaped density function with a
negative region in the center (gray), a signature of the
state's quantumness \cite{kenfack_JOB_wigner_nonclassicality}.
Experimentally, the resonator can be prepared in a similar,
albeit classical, state by further increasing the pump strength
[Fig. \ref{fig:nongaussian}\sub{c}]. This creates a pdf where
the resonator is preferentially located on a circular region
with an amplitude of about 30 pm. There is a clear reduction of
the probability of finding the resonator near the origin. This
can be understood as follows: When the resonator has a small
amplitude, comparable to the imprecision noise, the phase
estimation is not very reliable and the squeezing is
ineffective, pushing the resonator away from the origin.
However, as the amplitude grows due the error $\hat \varphi-
\varphi$, the phase estimation gets more accurate and the
resonator gets squeezed back toward the origin until a dynamic
equilibrium is reached. It is thus less likely to find the
resonator at small amplitudes, reducing the pdf near the
origin. A dip in the probability density is clearly visible in
the data. Note that self-sustained and parametric oscillations
would also show ring-like pdfs. However, this is not the case
here. First of all, the amplitude is much smaller than for
parametric oscillations; for $100 \un{mV}$ pump strength the
most likely amplitude 16 pm as indicated by the cuts in Fig.
\ref{fig:nongaussian}\sub{c}. This is two orders of magnitude
smaller than the much larger parametric oscillation amplitude
(typical a few nm) that result above threshold without
feedback. Furthermore, free oscillations are also ruled out
theoretically since Eq. (\ref{eq:Adotfeedback}) shows that
there is no fixed point except for the zero-amplitude solution
$A = 0$. Both the pdf and the cuts clearly show the deviation
from the Gaussian shape that one would obtain for regular
parametric squeezing or with linear active feedback cooling.
The nonlinearity of the feedback thus naturally leads to
non-Gaussian pdfs and this bring interesting prospects for
generating non-Gaussian quantum states.

The  symmetry of the pdf in Fig. \ref{fig:nongaussian}\sub{b}
indicates that the state of the resonator is phase-insensitive.
This is because the feedback squeezes isotropically as
explained above. However, it is also possible to generate
phase-dependent states where the circular symmetry is broken.
By applying a separate 2f pump signal with a \emph{fixed} phase
and yet keeping the co-rotating feedback on, a time dependence
is introduced and non-cylindrical pdfs should emerge. Figure
\ref{fig:nongaussian}\sub{e} shows that this is indeed the
case. For this combination of fixed and feedback pumps the
resonator is in an equal probability of two displaced coherent
states, analogous to the cat state shown in \sub{d}. This pdf
is the result of the feedback that forces the resonator to the
donut-like pdf of Fig. \ref{fig:nongaussian}(b) and the fixed
pump that pushes the resonator to the sigar-like shapes shown
in Fig. \ref{fig:parametric}. The resonator can thus be in two
different positions in phase space separated by a region of low
probabilitiy near the origin. The resonator dynamically
switches between them \cite{chan_PRE_parametric_switching}
resulting in the pdf that resembles the cat state in Fig.
\ref{fig:nongaussian}(d). Figure \ref{fig:nongaussian}\sub{f}
shows this switching, which is characterized by
telegraph-noise-like transitions from one displaced state to
the other. Although the feedback reduces the phase space of the
resonator to one of the two states for most of the time,
sometimes the thermal noise brings the resonator close to the
origin where the phase estimation is inaccurate. This way the
pump phase might become misaligned allowing the resonator to
switch over to the state with the opposite phase. Then it will
stay in that region of phase space until the next jump occurs.

\section{Conclusions and outlook}
We have demonstrated that our parametric feedback cannot only
be used to squeeze the thermomechanical motion by more than 3
dB but will also be an indispensable tool to prepare mechanical
resonators in highly nonlinear classical states. We emphasize
that everything demonstrated here can in principle be extended
to the quantum regime \cite{poot_physrep_quantum_regime}: The
parametric actuation can be done using
quantum-backaction-limited optical fields
\cite{purdy_science_RPSN} through the optical spring effect,
and also the optical readout is in principle shot-noise
limited. The nonlinear feedback will thus create correlations
between the quantum imprecision noise and the resonator motion.
Currently we have demonstrated our feedback scheme using
classical resonators but when employing cooling GHz resonators
close to their groundstate (either using a dilution
refrigerator \cite{oconnell_nature_quantum_piezo_resonator} or
using optical backaction cooling
\cite{chan_nature_groundstate}) the thermal motion vanishes and
is replaced by the zero-point motion allowing to employ our
scheme with a true quantum system. Then, in principle only
quantum noise will be fed back to the system and hence it seems
feasible that not only highly nonlinear states, but even true
non-classical mechanical states can be generated.

\begin{acknowledgements}
We thank Aashish Clerk for discussions. This work was partly
funded by the DARPA/MTO ORCHID program through a AFOSR grant.
M.P acknowledges the Netherlands Organization for Scientific
Research (NWO) / Marie Curie Cofund Action for a Rubicon
fellowship. H.X.T. thanks the Packard Foundations for a
Fellowship in Science and Engineering and the National Science
Foundation for a career award. Facilities used were supported
by Yale Institute for Nanoscience and Quantum Engineering and
NSF MRSEC DMR 1119826.
\end{acknowledgements}


\begin{thebibliography}{42}
\expandafter\ifx\csname
natexlab\endcsname\relax\def\natexlab#1{#1}\fi
\expandafter\ifx\csname bibnamefont\endcsname\relax
  \def\bibnamefont#1{#1}\fi
\expandafter\ifx\csname bibfnamefont\endcsname\relax
  \def\bibfnamefont#1{#1}\fi
\expandafter\ifx\csname citenamefont\endcsname\relax
  \def\citenamefont#1{#1}\fi
\expandafter\ifx\csname url\endcsname\relax
  \def\url#1{\texttt{#1}}\fi
\expandafter\ifx\csname
urlprefix\endcsname\relax\def\urlprefix{URL }\fi
\providecommand{\bibinfo}[2]{#2}
\providecommand{\eprint}[2][]{\url{#2}}

\bibitem[{\citenamefont{Poot and van~der
  Zant}(2012)}]{poot_physrep_quantum_regime}
\bibinfo{author}{\bibfnamefont{M.}~\bibnamefont{Poot}}
\bibnamefont{and}
  \bibinfo{author}{\bibfnamefont{H.~S.~J.} \bibnamefont{van~der Zant}},
  \href{http://www.sciencedirect.com/science/article/pii/S0370157311003644}
  {\emph{\bibinfo{title}{Mechanical systems in the quantum regime}}},
  \bibinfo{journal}{Phys. Rep.} \textbf{\bibinfo{volume}{511}},
  \bibinfo{pages}{273} (\bibinfo{year}{2012}).

\bibitem[{\citenamefont{O'Connell
    et~al.}(2010)\citenamefont{O'Connell,
  Hofheinz, Ansmann, Bialczak, Lenander, Lucero, Neeley, Sank, Wang, Weides
  et~al.}}]{oconnell_nature_quantum_piezo_resonator}
\bibinfo{author}{\bibfnamefont{A.~D.} \bibnamefont{O'Connell}},
  \bibinfo{author}{\bibfnamefont{M.}~\bibnamefont{Hofheinz}},
  \bibinfo{author}{\bibfnamefont{M.}~\bibnamefont{Ansmann}},
  \bibinfo{author}{\bibfnamefont{R.~C.} \bibnamefont{Bialczak}},
  \bibinfo{author}{\bibfnamefont{M.}~\bibnamefont{Lenander}},
  \bibinfo{author}{\bibfnamefont{E.}~\bibnamefont{Lucero}},
  \bibinfo{author}{\bibfnamefont{M.}~\bibnamefont{Neeley}},
  \bibinfo{author}{\bibfnamefont{D.}~\bibnamefont{Sank}},
  \bibinfo{author}{\bibfnamefont{H.}~\bibnamefont{Wang}},
  \bibinfo{author}{\bibfnamefont{M.}~\bibnamefont{Weides}},
  \bibnamefont{et~al.}, \href{http://dx.doi.org/10.1038/nature08967}
  {\emph{\bibinfo{title}{Quantum ground state and single-phonon control of a
  mechanical resonator}}}, \bibinfo{journal}{Nature}
  \textbf{\bibinfo{volume}{464}}, \bibinfo{pages}{697} (\bibinfo{year}{2010}).

\bibitem[{\citenamefont{Chan et~al.}(2011)\citenamefont{Chan,
    Alegre,
  Safavi-Naeini, Hill, Krause, Groblacher, Aspelmeyer, and
  Painter}}]{chan_nature_groundstate}
\bibinfo{author}{\bibfnamefont{J.}~\bibnamefont{Chan}},
  \bibinfo{author}{\bibfnamefont{T.~P.~M.} \bibnamefont{Alegre}},
  \bibinfo{author}{\bibfnamefont{A.~H.} \bibnamefont{Safavi-Naeini}},
  \bibinfo{author}{\bibfnamefont{J.~T.} \bibnamefont{Hill}},
  \bibinfo{author}{\bibfnamefont{A.}~\bibnamefont{Krause}},
  \bibinfo{author}{\bibfnamefont{S.}~\bibnamefont{Groblacher}},
  \bibinfo{author}{\bibfnamefont{M.}~\bibnamefont{Aspelmeyer}},
  \bibnamefont{and} \bibinfo{author}{\bibfnamefont{O.}~\bibnamefont{Painter}},
  \href{http://dx.doi.org/10.1038/nature10461} {\emph{\bibinfo{title}{Laser
  cooling of a nanomechanical oscillator into its quantum ground state}}},
  \bibinfo{journal}{Nature} \textbf{\bibinfo{volume}{478}}, \bibinfo{pages}{89}
  (\bibinfo{year}{2011}).

\bibitem[{\citenamefont{Teufel
    et~al.}({2011})\citenamefont{Teufel, Donner, Li,
  Harlow, Allman, Cicak, Sirois, Whittaker, Lehnert, and
  Simmonds}}]{teufel_nature_groundstate}
\bibinfo{author}{\bibfnamefont{J.~D.} \bibnamefont{Teufel}},
  \bibinfo{author}{\bibfnamefont{T.}~\bibnamefont{Donner}},
  \bibinfo{author}{\bibfnamefont{D.}~\bibnamefont{Li}},
  \bibinfo{author}{\bibfnamefont{J.~W.} \bibnamefont{Harlow}},
  \bibinfo{author}{\bibfnamefont{M.~S.} \bibnamefont{Allman}},
  \bibinfo{author}{\bibfnamefont{K.}~\bibnamefont{Cicak}},
  \bibinfo{author}{\bibfnamefont{A.~J.} \bibnamefont{Sirois}},
  \bibinfo{author}{\bibfnamefont{J.~D.} \bibnamefont{Whittaker}},
  \bibinfo{author}{\bibfnamefont{K.~W.} \bibnamefont{Lehnert}},
  \bibnamefont{and} \bibinfo{author}{\bibfnamefont{R.~W.}
  \bibnamefont{Simmonds}}, \href{http://dx.doi.org/{10.1038/nature10261}}
  {\emph{\bibinfo{title}{{Sideband cooling of micromechanical motion to the
  quantum ground state}}}}, \bibinfo{journal}{Nature}
  \textbf{\bibinfo{volume}{{475}}}, \bibinfo{pages}{359}
  (\bibinfo{year}{{2011}}).

\bibitem[{\citenamefont{Gr\"oblacher
    et~al.}(2009)\citenamefont{Gr\"oblacher,
  Hammerer, Vanner, and Aspelmeyer}}]{groeblacher_nature_strong_coupling}
\bibinfo{author}{\bibfnamefont{S.}~\bibnamefont{Gr\"oblacher}},
  \bibinfo{author}{\bibfnamefont{K.}~\bibnamefont{Hammerer}},
  \bibinfo{author}{\bibfnamefont{M.~R.} \bibnamefont{Vanner}},
  \bibnamefont{and}
  \bibinfo{author}{\bibfnamefont{M.}~\bibnamefont{Aspelmeyer}},
  \href{http://dx.doi.org/10.1038/nature08171}
  {\emph{\bibinfo{title}{Observation of strong coupling between a
  micromechanical resonator and an optical cavity field}}},
  \bibinfo{journal}{Nature} \textbf{\bibinfo{volume}{460}},
  \bibinfo{pages}{724} (\bibinfo{year}{2009}).

\bibitem[{\citenamefont{Weis et~al.}(2010)\citenamefont{Weis,
    Rivière,
  Deléglise, Gavartin, Arcizet, Schliesser, and
  Kippenberg}}]{weis_science_transparency}
\bibinfo{author}{\bibfnamefont{S.}~\bibnamefont{Weis}},
  \bibinfo{author}{\bibfnamefont{R.}~\bibnamefont{Rivière}},
  \bibinfo{author}{\bibfnamefont{S.}~\bibnamefont{Deléglise}},
  \bibinfo{author}{\bibfnamefont{E.}~\bibnamefont{Gavartin}},
  \bibinfo{author}{\bibfnamefont{O.}~\bibnamefont{Arcizet}},
  \bibinfo{author}{\bibfnamefont{A.}~\bibnamefont{Schliesser}},
  \bibnamefont{and} \bibinfo{author}{\bibfnamefont{T.~J.}
  \bibnamefont{Kippenberg}},
  \href{http://www.sciencemag.org/content/330/6010/1520.abstract}
  {\emph{\bibinfo{title}{Optomechanically Induced Transparency}}},
  \bibinfo{journal}{Science} \textbf{\bibinfo{volume}{330}},
  \bibinfo{pages}{1520} (\bibinfo{year}{2010}).

\bibitem[{\citenamefont{Teufel
    et~al.}(2011)\citenamefont{Teufel, Li, Allman,
  Cicak, Sirois, Whittaker, and Simmonds}}]{teufel_nature_strong_coupling}
\bibinfo{author}{\bibfnamefont{J.~D.} \bibnamefont{Teufel}},
  \bibinfo{author}{\bibfnamefont{D.}~\bibnamefont{Li}},
  \bibinfo{author}{\bibfnamefont{M.~S.} \bibnamefont{Allman}},
  \bibinfo{author}{\bibfnamefont{K.}~\bibnamefont{Cicak}},
  \bibinfo{author}{\bibfnamefont{A.~J.} \bibnamefont{Sirois}},
  \bibinfo{author}{\bibfnamefont{J.~D.} \bibnamefont{Whittaker}},
  \bibnamefont{and} \bibinfo{author}{\bibfnamefont{R.~W.}
  \bibnamefont{Simmonds}}, \href{http://dx.doi.org/10.1038/nature09898}
  {\emph{\bibinfo{title}{Circuit cavity electromechanics in the strong-coupling
  regime}}}, \bibinfo{journal}{Nature} \textbf{\bibinfo{volume}{471}},
  \bibinfo{pages}{204} (\bibinfo{year}{2011}).

\bibitem[{\citenamefont{Palomaki
  et~al.}(2013{\natexlab{a}})\citenamefont{Palomaki, Teufel, Simmonds, and
  Lehnert}}]{palomaki_science_entanglement}
\bibinfo{author}{\bibfnamefont{T.~A.} \bibnamefont{Palomaki}},
  \bibinfo{author}{\bibfnamefont{J.~D.} \bibnamefont{Teufel}},
  \bibinfo{author}{\bibfnamefont{R.~W.} \bibnamefont{Simmonds}},
  \bibnamefont{and} \bibinfo{author}{\bibfnamefont{K.~W.}
  \bibnamefont{Lehnert}},
  \href{http://www.sciencemag.org/content/342/6159/710.abstract}
  {\emph{\bibinfo{title}{Entangling Mechanical Motion with Microwave Fields}}},
  \bibinfo{journal}{Science} \textbf{\bibinfo{volume}{342}},
  \bibinfo{pages}{710} (\bibinfo{year}{2013}{\natexlab{a}}).

\bibitem[{\citenamefont{Brooks
    et~al.}(2012)\citenamefont{Brooks, Botter,
  Schreppler, Purdy, Brahms, and
  Stamper-Kurn}}]{brooks_nature_atom_motion_light_squeezing}
\bibinfo{author}{\bibfnamefont{D.~W.} \bibnamefont{Brooks}},
  \bibinfo{author}{\bibfnamefont{T.}~\bibnamefont{Botter}},
  \bibinfo{author}{\bibfnamefont{S.}~\bibnamefont{Schreppler}},
  \bibinfo{author}{\bibfnamefont{T.~P.} \bibnamefont{Purdy}},
  \bibinfo{author}{\bibfnamefont{N.}~\bibnamefont{Brahms}}, \bibnamefont{and}
  \bibinfo{author}{\bibfnamefont{D.~M.} \bibnamefont{Stamper-Kurn}},
  \emph{\bibinfo{title}{Non-classical light generated by quantum-noise-driven
  cavity optomechanics}}, \bibinfo{journal}{Nature}
  \textbf{\bibinfo{volume}{488}}, \bibinfo{pages}{476} (\bibinfo{year}{2012}).

\bibitem[{\citenamefont{Safavi-Naeini
    et~al.}(2013)\citenamefont{Safavi-Naeini,
  Groblacher, Hill, Chan, Aspelmeyer, and
  Painter}}]{safavi-naeini_nature_optomechanical_squeezing}
\bibinfo{author}{\bibfnamefont{A.~H.}
\bibnamefont{Safavi-Naeini}},
  \bibinfo{author}{\bibfnamefont{S.}~\bibnamefont{Groblacher}},
  \bibinfo{author}{\bibfnamefont{J.~T.} \bibnamefont{Hill}},
  \bibinfo{author}{\bibfnamefont{J.}~\bibnamefont{Chan}},
  \bibinfo{author}{\bibfnamefont{M.}~\bibnamefont{Aspelmeyer}},
  \bibnamefont{and} \bibinfo{author}{\bibfnamefont{O.}~\bibnamefont{Painter}},
  \href{http://dx.doi.org/10.1038/nature12307} {\emph{\bibinfo{title}{Squeezed
  light from a silicon micromechanical resonator}}}, \bibinfo{journal}{Nature}
  \textbf{\bibinfo{volume}{500}}, \bibinfo{pages}{185} (\bibinfo{year}{2013}).

\bibitem[{\citenamefont{Purdy
    et~al.}(2013{\natexlab{a}})\citenamefont{Purdy,
  Yu, Peterson, Kampel, and Regal}}]{purdy_PRX_optomechanical_squeezing}
\bibinfo{author}{\bibfnamefont{T.~P.} \bibnamefont{Purdy}},
  \bibinfo{author}{\bibfnamefont{P.-L.} \bibnamefont{Yu}},
  \bibinfo{author}{\bibfnamefont{R.~W.} \bibnamefont{Peterson}},
  \bibinfo{author}{\bibfnamefont{N.~S.} \bibnamefont{Kampel}},
  \bibnamefont{and} \bibinfo{author}{\bibfnamefont{C.~A.} \bibnamefont{Regal}},
  \href{http://link.aps.org/doi/10.1103/PhysRevX.3.031012}
  {\emph{\bibinfo{title}{Strong Optomechanical Squeezing of Light}}},
  \bibinfo{journal}{Phys. Rev. X} \textbf{\bibinfo{volume}{3}},
  \bibinfo{pages}{031012} (\bibinfo{year}{2013}{\natexlab{a}}).

\bibitem[{\citenamefont{Palomaki
  et~al.}(2013{\natexlab{b}})\citenamefont{Palomaki, Harlow, Teufel, Simmonds,
  and Lehnert}}]{palomaki_nature_coherent_transfer}
\bibinfo{author}{\bibfnamefont{T.~A.} \bibnamefont{Palomaki}},
  \bibinfo{author}{\bibfnamefont{J.~W.} \bibnamefont{Harlow}},
  \bibinfo{author}{\bibfnamefont{J.~D.} \bibnamefont{Teufel}},
  \bibinfo{author}{\bibfnamefont{R.~W.} \bibnamefont{Simmonds}},
  \bibnamefont{and} \bibinfo{author}{\bibfnamefont{K.~W.}
  \bibnamefont{Lehnert}}, \href{http://dx.doi.org/10.1038/nature11915}
  {\emph{\bibinfo{title}{Coherent state transfer between itinerant microwave
  fields and a mechanical oscillator}}}, \bibinfo{journal}{Nature}
  \textbf{\bibinfo{volume}{495}}, \bibinfo{pages}{210}
  (\bibinfo{year}{2013}{\natexlab{b}}).

\bibitem[{\citenamefont{Okamoto
    et~al.}(2013)\citenamefont{Okamoto, Gourgout,
  Chang, Onomitsu, Mahboob, Chang, and Yamaguchi}}]{okamoto_natphys_coupled}
\bibinfo{author}{\bibfnamefont{H.}~\bibnamefont{Okamoto}},
  \bibinfo{author}{\bibfnamefont{A.}~\bibnamefont{Gourgout}},
  \bibinfo{author}{\bibfnamefont{C.-Y.} \bibnamefont{Chang}},
  \bibinfo{author}{\bibfnamefont{K.}~\bibnamefont{Onomitsu}},
  \bibinfo{author}{\bibfnamefont{I.}~\bibnamefont{Mahboob}},
  \bibinfo{author}{\bibfnamefont{E.~Y.} \bibnamefont{Chang}}, \bibnamefont{and}
  \bibinfo{author}{\bibfnamefont{H.}~\bibnamefont{Yamaguchi}},
  \href{http://dx.doi.org/10.1038/nphys2665} {\emph{\bibinfo{title}{Coherent
  phonon manipulation in coupled mechanical resonators}}},
  \bibinfo{journal}{Nat Phys} \textbf{\bibinfo{volume}{9}},
  \bibinfo{pages}{480} (\bibinfo{year}{2013}).

\bibitem[{\citenamefont{Faust et~al.}(2013)\citenamefont{Faust,
    Rieger,
  Seitner, Kotthaus, and Weig}}]{faust_natphys_coherent_control}
\bibinfo{author}{\bibfnamefont{T.}~\bibnamefont{Faust}},
  \bibinfo{author}{\bibfnamefont{J.}~\bibnamefont{Rieger}},
  \bibinfo{author}{\bibfnamefont{M.~J.} \bibnamefont{Seitner}},
  \bibinfo{author}{\bibfnamefont{J.~P.} \bibnamefont{Kotthaus}},
  \bibnamefont{and} \bibinfo{author}{\bibfnamefont{E.~M.} \bibnamefont{Weig}},
  \href{http://dx.doi.org/10.1038/nphys2666} {\emph{\bibinfo{title}{Coherent
  control of a classical nanomechanical two-level system}}},
  \bibinfo{journal}{Nat Phys} \textbf{\bibinfo{volume}{9}},
  \bibinfo{pages}{485} (\bibinfo{year}{2013}).

\bibitem[{\citenamefont{Bose et~al.}(1997)\citenamefont{Bose,
    Jacobs, and
  Knight}}]{bose_PRA_nonclassical_preparation}
\bibinfo{author}{\bibfnamefont{S.}~\bibnamefont{Bose}},
  \bibinfo{author}{\bibfnamefont{K.}~\bibnamefont{Jacobs}}, \bibnamefont{and}
  \bibinfo{author}{\bibfnamefont{P.~L.} \bibnamefont{Knight}},
  \href{http://link.aps.org/doi/10.1103/PhysRevA.56.4175}
  {\emph{\bibinfo{title}{Preparation of nonclassical states in cavities with a
  moving mirror}}}, \bibinfo{journal}{Phys. Rev. A}
  \textbf{\bibinfo{volume}{56}}, \bibinfo{pages}{4175} (\bibinfo{year}{1997}).

\bibitem[{\citenamefont{Vanner
    et~al.}(2011)\citenamefont{Vanner, Pikovski,
  Cole, Kim, Brukner, Hammerer, Milburn, and
  Aspelmeyer}}]{vanner_PNAS_pulsed_optomechanics}
\bibinfo{author}{\bibfnamefont{M.~R.} \bibnamefont{Vanner}},
  \bibinfo{author}{\bibfnamefont{I.}~\bibnamefont{Pikovski}},
  \bibinfo{author}{\bibfnamefont{G.~D.} \bibnamefont{Cole}},
  \bibinfo{author}{\bibfnamefont{M.~S.} \bibnamefont{Kim}},
  \bibinfo{author}{\bibfnamefont{u.}~\bibnamefont{Brukner}},
  \bibinfo{author}{\bibfnamefont{K.}~\bibnamefont{Hammerer}},
  \bibinfo{author}{\bibfnamefont{G.~J.} \bibnamefont{Milburn}},
  \bibnamefont{and}
  \bibinfo{author}{\bibfnamefont{M.}~\bibnamefont{Aspelmeyer}},
  \href{http://www.pnas.org/content/108/39/16182.abstract}
  {\emph{\bibinfo{title}{Pulsed quantum optomechanics}}},
  \bibinfo{journal}{Proceedings of the National Academy of Sciences}
  \textbf{\bibinfo{volume}{108}}, \bibinfo{pages}{16182}
  (\bibinfo{year}{2011}).

\bibitem[{\citenamefont{Vanner}(2011)}]{vanner_PRX_linear_quadratic}
    \bibinfo{author}{\bibfnamefont{M.~R.}
    \bibnamefont{Vanner}},
  \href{http://link.aps.org/doi/10.1103/PhysRevX.1.021011}
  {\emph{\bibinfo{title}{Selective Linear or Quadratic Optomechanical Coupling
  via Measurement}}}, \bibinfo{journal}{Phys. Rev. X}
  \textbf{\bibinfo{volume}{1}}, \bibinfo{pages}{021011} (\bibinfo{year}{2011}).

\bibitem[{\citenamefont{Romero-Isart
    et~al.}(2011)\citenamefont{Romero-Isart,
  Pflanzer, Blaser, Kaltenbaek, Kiesel, Aspelmeyer, and
  Cirac}}]{romero-isart_PRL_superpostions}
\bibinfo{author}{\bibfnamefont{O.}~\bibnamefont{Romero-Isart}},
  \bibinfo{author}{\bibfnamefont{A.~C.} \bibnamefont{Pflanzer}},
  \bibinfo{author}{\bibfnamefont{F.}~\bibnamefont{Blaser}},
  \bibinfo{author}{\bibfnamefont{R.}~\bibnamefont{Kaltenbaek}},
  \bibinfo{author}{\bibfnamefont{N.}~\bibnamefont{Kiesel}},
  \bibinfo{author}{\bibfnamefont{M.}~\bibnamefont{Aspelmeyer}},
  \bibnamefont{and} \bibinfo{author}{\bibfnamefont{J.~I.} \bibnamefont{Cirac}},
  \href{http://link.aps.org/doi/10.1103/PhysRevLett.107.020405}
  {\emph{\bibinfo{title}{Large Quantum Superpositions and Interference of
  Massive Nanometer-Sized Objects}}}, \bibinfo{journal}{Phys. Rev. Lett.}
  \textbf{\bibinfo{volume}{107}}, \bibinfo{pages}{020405}
  (\bibinfo{year}{2011}).

\bibitem[{\citenamefont{Clerk et~al.}(2008)\citenamefont{Clerk,
    Marquardt, and
  Jacobs}}]{clerk_NJP_BAE_squeezing}
\bibinfo{author}{\bibfnamefont{A.~A.} \bibnamefont{Clerk}},
  \bibinfo{author}{\bibfnamefont{F.}~\bibnamefont{Marquardt}},
  \bibnamefont{and} \bibinfo{author}{\bibfnamefont{K.}~\bibnamefont{Jacobs}},
  \href{http://stacks.iop.org/1367-2630/10/i=9/a=095010}
  {\emph{\bibinfo{title}{Back-action evasion and squeezing of a mechanical
  resonator using a cavity detector}}}, \bibinfo{journal}{New Journal of
  Physics} \textbf{\bibinfo{volume}{10}}, \bibinfo{pages}{095010}
  (\bibinfo{year}{2008}).

\bibitem[{\citenamefont{Rugar and
    Gr\"utter}(1991)}]{rugar_PRL_squeezing}
    \bibinfo{author}{\bibfnamefont{D.}~\bibnamefont{Rugar}}
    \bibnamefont{and}
  \bibinfo{author}{\bibfnamefont{P.}~\bibnamefont{Gr\"utter}},
  \href{http://dx.doi.org/10.1103/PhysRevLett.67.699}
  {\emph{\bibinfo{title}{Mechanical parametric amplification and
  thermomechanical noise squeezing}}}, \bibinfo{journal}{Phys. Rev. Lett.}
  \textbf{\bibinfo{volume}{67}}, \bibinfo{pages}{699} (\bibinfo{year}{1991}).

\bibitem[{\citenamefont{Vinante and
  Falferi}(2013)}]{vinante_PRL_feedback_squeezing}
\bibinfo{author}{\bibfnamefont{A.}~\bibnamefont{Vinante}}
\bibnamefont{and}
  \bibinfo{author}{\bibfnamefont{P.}~\bibnamefont{Falferi}},
  \href{http://link.aps.org/doi/10.1103/PhysRevLett.111.207203}
  {\emph{\bibinfo{title}{Feedback-Enhanced Parametric Squeezing of Mechanical
  Motion}}}, \bibinfo{journal}{Phys. Rev. Lett.}
  \textbf{\bibinfo{volume}{111}}, \bibinfo{pages}{207203}
  (\bibinfo{year}{2013}).

\bibitem[{\citenamefont{Szorkovszky
    et~al.}(2011)\citenamefont{Szorkovszky,
  Doherty, Harris, and Bowen}}]{szorkovszky_PRL_squeezing_weak_measurement}
\bibinfo{author}{\bibfnamefont{A.}~\bibnamefont{Szorkovszky}},
  \bibinfo{author}{\bibfnamefont{A.~C.} \bibnamefont{Doherty}},
  \bibinfo{author}{\bibfnamefont{G.~I.} \bibnamefont{Harris}},
  \bibnamefont{and} \bibinfo{author}{\bibfnamefont{W.~P.} \bibnamefont{Bowen}},
  \href{http://link.aps.org/doi/10.1103/PhysRevLett.107.213603}
  {\emph{\bibinfo{title}{Mechanical Squeezing via Parametric Amplification and
  Weak Measurement}}}, \bibinfo{journal}{Phys. Rev. Lett.}
  \textbf{\bibinfo{volume}{107}}, \bibinfo{pages}{213603}
  (\bibinfo{year}{2011}).

\bibitem[{\citenamefont{Poot and
    Tang}(2014)}]{poot_apl_phaseshifter}
    \bibinfo{author}{\bibfnamefont{M.}~\bibnamefont{Poot}}
    \bibnamefont{and}
  \bibinfo{author}{\bibfnamefont{H.~X.} \bibnamefont{Tang}},
  \href{http://scitation.aip.org/content/aip/journal/apl/104/6/10.1063/1.4864257}
  {\emph{\bibinfo{title}{Broadband nanoelectromechanical phase shifting of
  light on a chip}}}, \bibinfo{journal}{Appl. Phys. Lett.}
  \textbf{\bibinfo{volume}{104}}, \bibinfo{eid}{061101} (\bibinfo{year}{2014}).

\bibitem[{\citenamefont{Poot and
    Tang}()}]{poot_H_electrostatic}
    \bibinfo{author}{\bibfnamefont{M.}~\bibnamefont{Poot}}
    \bibnamefont{and}
  \bibinfo{author}{\bibfnamefont{H.}~\bibnamefont{Tang}},
  \emph{\bibinfo{title}{Strong electrostatic coupling in an optomechanical
  resonator}}, \bibinfo{howpublished}{In preparation}.

\bibitem[{\citenamefont{Unterreithmeier
  et~al.}(2009)\citenamefont{Unterreithmeier, Weig, and
  Kotthaus}}]{unterreithmeier_nature_dielectric}
\bibinfo{author}{\bibfnamefont{Q.~P.}
\bibnamefont{Unterreithmeier}},
  \bibinfo{author}{\bibfnamefont{E.~M.} \bibnamefont{Weig}}, \bibnamefont{and}
  \bibinfo{author}{\bibfnamefont{J.~P.} \bibnamefont{Kotthaus}},
  \href{http://dx.doi.org/10.1038/nature07932} {\emph{\bibinfo{title}{Universal
  transduction scheme for nanomechanical systems based on dielectric forces}}},
  \bibinfo{journal}{Nature} \textbf{\bibinfo{volume}{458}},
  \bibinfo{pages}{1001} (\bibinfo{year}{2009}).

\bibitem[{\citenamefont{Kozinsky
    et~al.}(2006)\citenamefont{Kozinsky, Postma,
  Bargatin, and Roukes}}]{kozinsky_APL_dynamicrange_tuning}
\bibinfo{author}{\bibfnamefont{I.}~\bibnamefont{Kozinsky}},
  \bibinfo{author}{\bibfnamefont{H.~W.~C.} \bibnamefont{Postma}},
  \bibinfo{author}{\bibfnamefont{I.}~\bibnamefont{Bargatin}}, \bibnamefont{and}
  \bibinfo{author}{\bibfnamefont{M.~L.} \bibnamefont{Roukes}},
  \href{http://link.aip.org/link/?APL/88/253101/1}
  {\emph{\bibinfo{title}{Tuning nonlinearity, dynamic range, and frequency of
  nanomechanical resonators}}}, \bibinfo{journal}{Appl. Phys. Lett.}
  \textbf{\bibinfo{volume}{88}}, \bibinfo{eid}{253101}
  (pages~\bibinfo{numpages}{3}) (\bibinfo{year}{2006}).

\bibitem[{\citenamefont{Poot et~al.}(2012)\citenamefont{Poot,
    Fong, Bagheri,
  Pernice, and Tang}}]{poot_PRA_backaction_limits}
\bibinfo{author}{\bibfnamefont{M.}~\bibnamefont{Poot}},
  \bibinfo{author}{\bibfnamefont{K.~Y.} \bibnamefont{Fong}},
  \bibinfo{author}{\bibfnamefont{M.}~\bibnamefont{Bagheri}},
  \bibinfo{author}{\bibfnamefont{W.~H.~P.} \bibnamefont{Pernice}},
  \bibnamefont{and} \bibinfo{author}{\bibfnamefont{H.~X.} \bibnamefont{Tang}},
  \href{http://link.aps.org/doi/10.1103/PhysRevA.86.053826}
  {\emph{\bibinfo{title}{Backaction limits on self-sustained optomechanical
  oscillations}}}, \bibinfo{journal}{Phys. Rev. A}
  \textbf{\bibinfo{volume}{86}}, \bibinfo{pages}{053826}
  (\bibinfo{year}{2012}).

\bibitem[{not()}]{note:pdf_resonstruction} \bibinfo{note}{The
    demodulated voltages are converted back to displacements
  using the transduction factor obtained from the area under the thermal noise
  spectra without the pump (cf. Fig. \ref{fig:overview}\sub{e}). The variances
  in Fig. \ref{fig:parametric} are the eigenvalues of the covariance matrix of
  $X_1(t)$ and $X_2(t)$ and thus include some (unsqueezed) imprecision noise.
  The amount of squeezing of the motion reported is thus a lower bound. The
  pdfs are kernel densities where a Gaussian kernel is added on every sampled
  point in the $X_1-X_2$ space, thus creating a smoothened estimator of the
  pdf.}

\bibitem[{\citenamefont{Lifshitz and
    Cross}(2008)}]{lifshitz_coupled_NL}
    \bibinfo{author}{\bibfnamefont{R.}~\bibnamefont{Lifshitz}}
    \bibnamefont{and}
  \bibinfo{author}{\bibfnamefont{M.~C.} \bibnamefont{Cross}},
  \emph{\bibinfo{title}{Reviews of Nonlinear Dynamics and Complexity: Volume 1
  (Annual Reviews of Nonlinear Dynamics and Complexity)}}
  (\bibinfo{publisher}{Wiley-VCH}, \bibinfo{year}{2008}),
  chap.~\bibinfo{chapter}{1}, ISBN \bibinfo{isbn}{3527407294}.

\bibitem[{\citenamefont{Karabalin
    et~al.}(2010)\citenamefont{Karabalin,
  Masmanidis, and Roukes}}]{karabalin_APL_parametric_VHF}
\bibinfo{author}{\bibfnamefont{R.~B.} \bibnamefont{Karabalin}},
  \bibinfo{author}{\bibfnamefont{S.~C.} \bibnamefont{Masmanidis}},
  \bibnamefont{and} \bibinfo{author}{\bibfnamefont{M.~L.}
  \bibnamefont{Roukes}},
  \href{http://scitation.aip.org/content/aip/journal/apl/97/18/10.1063/1.3505500}
  {\emph{\bibinfo{title}{Efficient parametric amplification in high and very
  high frequency piezoelectric nanoelectromechanical systems}}},
  \bibinfo{journal}{Appl. Phys. Lett.} \textbf{\bibinfo{volume}{97}},
  \bibinfo{eid}{183101} (\bibinfo{year}{2010}).

\bibitem[{\citenamefont{Wu and
  Zhong}(2011)}]{chung-chiang_APL_parametric_SWCNT}
\bibinfo{author}{\bibfnamefont{C.-C.} \bibnamefont{Wu}}
\bibnamefont{and}
  \bibinfo{author}{\bibfnamefont{Z.}~\bibnamefont{Zhong}},
  \href{http://scitation.aip.org/content/aip/journal/apl/99/8/10.1063/1.3627178}
  {\emph{\bibinfo{title}{Parametric amplification in single-walled carbon
  nanotube nanoelectromechanical resonators}}}, \bibinfo{journal}{Appl. Phys.
  Lett.} \textbf{\bibinfo{volume}{99}}, \bibinfo{eid}{083110}
  (\bibinfo{year}{2011}).

\bibitem[{\citenamefont{Mahboob and
    Yamaguchi}(2008)}]{mahboob_natnano_bit}
    \bibinfo{author}{\bibfnamefont{I.}~\bibnamefont{Mahboob}}
    \bibnamefont{and}
  \bibinfo{author}{\bibfnamefont{H.}~\bibnamefont{Yamaguchi}},
  \href{http://dx.doi.org/10.1038/nnano.2008.84} {\emph{\bibinfo{title}{Bit
  storage and bit flip operations in an electromechanical oscillator}}},
  \bibinfo{journal}{Nat Nano} \textbf{\bibinfo{volume}{3}},
  \bibinfo{pages}{275} (\bibinfo{year}{2008}).

\bibitem[{\citenamefont{Eichler
    et~al.}(2011)\citenamefont{Eichler, Chaste,
  Moser, and Bachtold}}]{eichler_NL_parametric_amplification_CNT}
\bibinfo{author}{\bibfnamefont{A.}~\bibnamefont{Eichler}},
  \bibinfo{author}{\bibfnamefont{J.}~\bibnamefont{Chaste}},
  \bibinfo{author}{\bibfnamefont{J.}~\bibnamefont{Moser}}, \bibnamefont{and}
  \bibinfo{author}{\bibfnamefont{A.}~\bibnamefont{Bachtold}},
  \href{http://pubs.acs.org/doi/abs/10.1021/nl200950d}
  {\emph{\bibinfo{title}{Parametric Amplification and Self-Oscillation in a
  Nanotube Mechanical Resonator}}}, \bibinfo{journal}{Nano Lett.}
  \textbf{\bibinfo{volume}{11}}, \bibinfo{pages}{2699} (\bibinfo{year}{2011}).

\bibitem[{\citenamefont{Blencowe and
  Wybourne}(2000)}]{blencowe_physb_squeezing}
\bibinfo{author}{\bibfnamefont{M.}~\bibnamefont{Blencowe}}
\bibnamefont{and}
  \bibinfo{author}{\bibfnamefont{M.}~\bibnamefont{Wybourne}},
  \href{http://www.sciencedirect.com/science/article/pii/S0921452699018621}
  {\emph{\bibinfo{title}{Quantum squeezing of mechanical motion for
  micron-sized cantilevers}}}, \bibinfo{journal}{Physica B: Condensed Matter}
  \textbf{\bibinfo{volume}{280}}, \bibinfo{pages}{555} (\bibinfo{year}{2000}).

\bibitem[{\citenamefont{Szorkovszky
    et~al.}(2012)\citenamefont{Szorkovszky,
  Doherty, Harris, and Bowen}}]{szorkovszky_NJP_squeezing_position_estimation}
\bibinfo{author}{\bibfnamefont{A.}~\bibnamefont{Szorkovszky}},
  \bibinfo{author}{\bibfnamefont{A.~C.} \bibnamefont{Doherty}},
  \bibinfo{author}{\bibfnamefont{G.~I.} \bibnamefont{Harris}},
  \bibnamefont{and} \bibinfo{author}{\bibfnamefont{W.~P.} \bibnamefont{Bowen}},
  \href{http://stacks.iop.org/1367-2630/14/i=9/a=095026}
  {\emph{\bibinfo{title}{Position estimation of a parametrically driven
  optomechanical system}}}, \bibinfo{journal}{New J. Phys.}
  \textbf{\bibinfo{volume}{14}}, \bibinfo{pages}{095026}
  (\bibinfo{year}{2012}).

\bibitem[{\citenamefont{Gieseler
    et~al.}(2012)\citenamefont{Gieseler, Deutsch,
  Quidant, and Novotny}}]{gieseler_PRL_nanoparticle}
\bibinfo{author}{\bibfnamefont{J.}~\bibnamefont{Gieseler}},
  \bibinfo{author}{\bibfnamefont{B.}~\bibnamefont{Deutsch}},
  \bibinfo{author}{\bibfnamefont{R.}~\bibnamefont{Quidant}}, \bibnamefont{and}
  \bibinfo{author}{\bibfnamefont{L.}~\bibnamefont{Novotny}},
  \href{http://link.aps.org/doi/10.1103/PhysRevLett.109.103603}
  {\emph{\bibinfo{title}{Subkelvin Parametric Feedback Cooling of a
  Laser-Trapped Nanoparticle}}}, \bibinfo{journal}{Phys. Rev. Lett.}
  \textbf{\bibinfo{volume}{109}}, \bibinfo{pages}{103603}
  (\bibinfo{year}{2012}).

\bibitem[{\citenamefont{Poot et~al.}(2011)\citenamefont{Poot,
    Etaki, Yamaguchi,
  and van~der Zant}}]{poot_APL_cooling}
\bibinfo{author}{\bibfnamefont{M.}~\bibnamefont{Poot}},
  \bibinfo{author}{\bibfnamefont{S.}~\bibnamefont{Etaki}},
  \bibinfo{author}{\bibfnamefont{H.}~\bibnamefont{Yamaguchi}},
  \bibnamefont{and} \bibinfo{author}{\bibfnamefont{H.~S.~J.}
  \bibnamefont{van~der Zant}}, \href{http://link.aip.org/link/?APL/99/013113/1}
  {\emph{\bibinfo{title}{Discrete-time quadrature feedback cooling of a
  radio-frequency mechanical resonator}}}, \bibinfo{journal}{Appl. Phys. Lett.}
  \textbf{\bibinfo{volume}{99}}, \bibinfo{eid}{013113}
  (pages~\bibinfo{numpages}{3}) (\bibinfo{year}{2011}).

\bibitem[{\citenamefont{Kleckner and
  Bouwmeester}(2006)}]{kleckner_nature_feedback}
\bibinfo{author}{\bibfnamefont{D.}~\bibnamefont{Kleckner}}
\bibnamefont{and}
  \bibinfo{author}{\bibfnamefont{D.}~\bibnamefont{Bouwmeester}},
  \href{http://dx.doi.org/10.1038/nature05231}
  {\emph{\bibinfo{title}{Sub-kelvin optical cooling of a micromechanical
  resonator}}}, \bibinfo{journal}{Nature} \textbf{\bibinfo{volume}{444}},
  \bibinfo{pages}{75} (\bibinfo{year}{2006}).

\bibitem[{\citenamefont{Poggio
    et~al.}(2007)\citenamefont{Poggio, Degen, Mamin,
  and Rugar}}]{poggio_PRL_feedback}
\bibinfo{author}{\bibfnamefont{M.}~\bibnamefont{Poggio}},
  \bibinfo{author}{\bibfnamefont{C.~L.} \bibnamefont{Degen}},
  \bibinfo{author}{\bibfnamefont{H.~J.} \bibnamefont{Mamin}}, \bibnamefont{and}
  \bibinfo{author}{\bibfnamefont{D.}~\bibnamefont{Rugar}},
  \href{http://link.aps.org/abstract/PRL/v99/e017201}
  {\emph{\bibinfo{title}{Feedback Cooling of a Cantilever's Fundamental Mode
  below 5 {mK}}}}, \bibinfo{journal}{Phys. Rev. Lett.}
  \textbf{\bibinfo{volume}{99}}, \bibinfo{eid}{017201}
  (pages~\bibinfo{numpages}{4}) (\bibinfo{year}{2007}).

\bibitem[{\citenamefont{Kenfack and {\.
  Z}yczkowski}(2004)}]{kenfack_JOB_wigner_nonclassicality}
\bibinfo{author}{\bibfnamefont{A.}~\bibnamefont{Kenfack}}
\bibnamefont{and}
  \bibinfo{author}{\bibfnamefont{K.}~\bibnamefont{{\. Z}yczkowski}},
  \href{http://stacks.iop.org/1464-4266/6/i=10/a=003}
  {\emph{\bibinfo{title}{Negativity of the{ Wigner} function as an indicator of
  non-classicality}}}, \bibinfo{journal}{Journal of Optics B: Quantum and
  Semiclassical Optics} \textbf{\bibinfo{volume}{6}}, \bibinfo{pages}{396}
  (\bibinfo{year}{2004}).

\bibitem[{\citenamefont{Chan et~al.}(2008)\citenamefont{Chan,
    Dykman, and
  Stambaugh}}]{chan_PRE_parametric_switching}
\bibinfo{author}{\bibfnamefont{H.~B.} \bibnamefont{Chan}},
  \bibinfo{author}{\bibfnamefont{M.~I.} \bibnamefont{Dykman}},
  \bibnamefont{and}
  \bibinfo{author}{\bibfnamefont{C.}~\bibnamefont{Stambaugh}},
  \href{http://link.aps.org/doi/10.1103/PhysRevE.78.051109}
  {\emph{\bibinfo{title}{Switching-path distribution in multidimensional
  systems}}}, \bibinfo{journal}{Phys. Rev. E} \textbf{\bibinfo{volume}{78}},
  \bibinfo{pages}{051109} (\bibinfo{year}{2008}).

\bibitem[{\citenamefont{Purdy
    et~al.}(2013{\natexlab{b}})\citenamefont{Purdy,
  Peterson, and Regal}}]{purdy_science_RPSN}
\bibinfo{author}{\bibfnamefont{T.~P.} \bibnamefont{Purdy}},
  \bibinfo{author}{\bibfnamefont{R.~W.} \bibnamefont{Peterson}},
  \bibnamefont{and} \bibinfo{author}{\bibfnamefont{C.~A.} \bibnamefont{Regal}},
  \href{http://www.sciencemag.org/content/339/6121/801.abstract}
  {\emph{\bibinfo{title}{Observation of Radiation Pressure Shot Noise on a
  Macroscopic Object}}}, \bibinfo{journal}{Science}
  \textbf{\bibinfo{volume}{339}}, \bibinfo{pages}{801}
  (\bibinfo{year}{2013}{\natexlab{b}}).

\end{thebibliography}

\end{document}